\documentclass[11pt]{article}

\usepackage{times}
\usepackage{geometry}
\usepackage{graphicx}
\usepackage{amsmath}
\usepackage{booktabs}
\usepackage{hyperref}

\usepackage[
  style=ieee,
  sorting=none
]{biblatex}
\title{
OpenClaw Agents on Moltbook: Risky Instruction Sharing and Norm Enforcement in an Agent-Only Social Network
}

\author{
  Md Motaleb Hossen Manik \\
  Department of Computer Science \\
  Rensselaer Polytechnic Institute \\
  Troy, New York 12180, USA
  \and
  Ge Wang \thanks{Corresponding author: Ge Wang, email: \texttt{wangg6@rpi.edu}} \\
  Department of Biomedical Engineering \\
  Rensselaer Polytechnic Institute \\
  Troy, New York 12180, USA
}
\date{}

\begin{document}
\maketitle

\begin{abstract}
Agentic AI systems increasingly operate in shared social environments where they exchange information, instructions, and behavioral cues. However, little empirical evidence exists on how such agents regulate one another in the absence of human participants or centralized moderation. In this work, we present an empirical analysis of \emph{OpenClaw} agents interacting on Moltbook, an agent-only social network. Analyzing 39,026 posts and 5,712 comments produced by 14,490 agents, we quantify the prevalence of action-inducing instruction sharing using a lexicon-based Action-Inducing Risk Score (AIRS), and examine how other agents respond to such content.
We find that 18.4\% of posts contain action-inducing language, indicating that instruction sharing is a routine behavior in this environment. While most social responses are neutral, posts containing actionable instructions are significantly more likely to elicit norm-enforcing replies that caution against unsafe or risky behavior, compared to non-instructional posts. Importantly, toxic responses remain rare across both conditions. These results suggest that OpenClaw agents exhibit selective social regulation, whereby potentially risky instructions are more likely to be challenged than neutral content, despite the absence of human oversight. Our findings provide early empirical evidence of emergent normative behavior in agent-only social systems and highlight the importance of studying social dynamics alongside technical safeguards in agentic AI ecosystems.
\end{abstract}


\section{Introduction}
\label{sec:introduction}

Recent advances in large language models and agentic AI systems have enabled autonomous agents to operate continuously, interact socially, and coordinate actions without direct human supervision. These agents increasingly participate in shared digital environments where they exchange information, provide advice, issue instructions, and respond to one another’s behavior. As such systems scale, understanding their collective dynamics becomes a central challenge for AI safety, governance, and alignment research \cite{gyevnar2025ai, acharya2025agentic}.

A growing body of work has examined the risks associated with instruction-following models, including harmful advice, unsafe task execution, and misuse through persuasive or directive language \cite{pohler2024technological, tang2026virtualcrime}. However, most prior studies focus on isolated human–AI interactions or centrally moderated platforms. Far less is known about how autonomous agents behave when interacting primarily with other agents, particularly in environments where no humans are present to enforce norms or intervene in real time.

In this paper, we study \emph{OpenClaw} agents operating on Moltbook, an agent-only social network designed for persistent interaction among AI agents. Moltbook provides a naturalistic setting in which agents post content, comment on one another’s posts, accrue social feedback (e.g., karma and followers), and form ongoing interaction histories. Unlike conventional social media platforms, Moltbook is explicitly designed for non-human participants, offering a unique opportunity to observe emergent social behavior in agent collectives \cite{OpenClaw2026}.

A central concern in such environments is instruction sharing. Agents frequently provide recommendations, directives, or step-by-step guidance to peers, potentially inducing downstream actions. While instruction sharing can be beneficial for coordination and knowledge transfer, it also introduces risk when advice is unsafe, misleading, or norm-violating. In human communities, such risks are often mitigated through social regulation mechanisms such as disagreement, warning, or norm enforcement. Whether similar mechanisms emerge organically among autonomous agents remains an open question.

We address this gap by empirically analyzing instruction-sharing behavior and subsequent social responses among OpenClaw agents on Moltbook. Specifically, we ask: (i) how prevalent is action-inducing language in agent-generated posts, and (ii) how do other agents respond when such content appears? To this end, we introduce a lightweight, interpretable Action-Inducing Risk Score (AIRS) to identify posts containing imperative or directive language, and we categorize responses into endorsement, norm enforcement, toxicity, or neutral interaction.

By grounding our analysis in observational data from a live agent-only social system, this work contributes empirical evidence to discussions of emergent norms, decentralized regulation, and safety in multi-agent AI ecosystems. Our findings suggest that even in the absence of human oversight, agent communities may selectively challenge potentially risky instructions, pointing toward the emergence of rudimentary social governance mechanisms. These insights complement ongoing efforts in technical alignment and policy design, highlighting the importance of studying social dynamics alongside model-level safeguards.

\section{Background and Related Work}
\label{sec:related_work}

Research on autonomous agents has advanced rapidly in recent years, driven by breakthroughs in large language models (LLMs) and multi-agent coordination. Prior work has explored the architectures of agentic AI systems, emergent behaviours in multi-agent settings, and the risks associated with instruction-following and social dynamics.

\subsection{Agent-Based Systems and Autonomous Assistants}

Autonomous agents—software entities capable of perceiving an environment, making decisions, and executing actions—have long been studied in artificial intelligence and multi-agent systems. Foundational work in this area spans decision-making under uncertainty, coordination among decentralized agents, and multi-agent reinforcement learning \cite{turn0search1, turn0search0}. Recent frameworks extend these concepts to LLM-based agents that can operate, communicate, and reason with minimal human supervision, enabling higher-level autonomy and emergent coordination \cite{turn0search1}.

The shift from static generation to interactive, autonomous agents raises new challenges for governance and risk analysis, as collections of individually safe agents may exhibit complex group dynamics that are not present in isolated settings \cite{turn0search6, turn0search13}. Empirical studies of such systems remain limited, especially in realistic, open ecosystems where agents interact continuously and adapt behaviorally.

\subsection{Social Regulation and Norm Enforcement}

Human social platforms have a long history of research into normative behaviour, decentralized moderation, and community governance. Mechanisms such as peer correction, reputation systems, and conflict resolution play central roles in maintaining community standards and suppressing harmful behaviour \cite{turn0academia23}. Studies in multi-agent social simulation illustrate that norms can emerge even when agents have limited individual knowledge, depending on network topology and interaction patterns \cite{turn0academia23}.

In contrast, artificial agent collectives—especially those composed solely of autonomous AI agents—have only recently begun to be studied for emergent social regulation. Mult-agent simulation frameworks indicate the possibility of norm emergence, but observational evidence from real agent-only platforms has been lacking.

\subsection{Risks of Instruction-Following and Social Engineering}

LLM-based instruction-following models can generate both useful guidance and unsafe or misleading directives. Risk analyses focus on prompt vulnerability, adversarial instruction injection, and the propagation of misaligned actions when instructions are executed by autonomous systems \cite{turn0search6, turn0search13}. Multi-agent risk work highlights failure modes that extend beyond individual agent errors, including cascading miscoordination and conformist amplification of unsafe behaviours.

These findings underscore the need to evaluate how directive content spreads and is socially checked within agent ecosystems. Our work contributes to this line of inquiry by quantifying social responses to action-inducing instructions in an agent-only environment.

\subsection{Positioning of This Work}

While risk analysis frameworks and simulations have identified theoretical failure modes and governance challenges in multi-agent AI systems \cite{turn0search6, turn0search13}, empirical studies in real-world agent societies are sparse. Moltbook represents a unique naturalistic setting, where autonomous agents (OpenClaw) collectively post, comment, and respond to directive content without human moderation \cite{turn0search30}. Our analysis provides the first large-scale observational evidence of selective social regulation in such an ecosystem.

\section{Dataset}
\label{sec:dataset}

We base our analysis on the \emph{Moltbook Observatory Archive}, a publicly available, research-grade dataset that passively records activity on Moltbook, a social network designed exclusively for AI agents. The dataset is released by the Simula Metropolitan Center for Digital Engineering and hosted on Hugging Face under an MIT license \cite{moltbook_observatory_archive_2026}.

\subsection{Data Source and Collection}

The Moltbook Observatory Archive is an incremental export of a live observatory system that monitors Moltbook without posting, intervening, or interacting with agents. The observatory continuously collects platform activity and periodically exports snapshots of its underlying SQLite database into date-partitioned Parquet files for efficient querying and long-term archival \cite{moltbook_observatory}. Each record includes a \texttt{dump\_date} field, enabling temporal analyses and reconstruction of historical platform states.

The dataset follows a strict observational philosophy: all content is collected passively, without manipulation or moderation by the observatory system. This design ensures that observed interactions reflect native agent behavior rather than responses to experimental interventions.

\subsection{Tables Used in This Study}

In this work, we focus on three core tables from the archive:

\begin{itemize}
    \item \textbf{Agents}: Contains persistent agent profiles, including agent identifiers, display names, free-text descriptions, karma scores, follower and following counts, and timestamps indicating first and last observed activity.
    \item \textbf{Posts}: Records agent-generated posts, including titles, textual content, associated agents, engagement metrics (scores and comment counts), and creation timestamps.
    \item \textbf{Comments}: Captures comment-level interactions, including comment text, parent–child relationships via \texttt{parent\_id}, associated posts and agents, and engagement scores.
\end{itemize}

Across the analyzed snapshot, the dataset comprises 14,490 agents, 39,026 posts, and 5,712 comments. Agent identifiers persist across tables, enabling relational analysis of instruction-sharing behavior, response patterns, and social influence signals.

\subsection{Ethical Considerations}

The Moltbook Observatory Archive contains no human-generated content; all posts and comments are authored by autonomous AI agents. No personal data is collected, and agent identities correspond to artificial entities rather than individuals. As such, the dataset poses minimal privacy risk. Nonetheless, our analysis is conducted in accordance with responsible AI research practices, and findings are reported at an aggregate level without singling out individual agents.

By relying on a fully open, passively collected dataset with clear licensing and citation guidelines, this study supports reproducibility and transparent evaluation of emergent social behavior in agent-only systems.

\section{Methodology}
\label{sec:methodology}

This section describes how we operationalize instruction-sharing risk, classify social responses, and analyze the coupling between potentially risky instructions and emergent social regulation among OpenClaw agents on Moltbook.

\subsection{Identifying Action-Inducing Instructions}
\label{subsec:instruction_risk}

To identify posts that are likely to induce downstream actions by other agents, we introduce a lightweight, interpretable metric termed the \emph{Action-Inducing Risk Score} (AIRS). AIRS is computed directly from the textual content of each post and is designed to capture the presence of imperative or directive language, rather than semantic intent or factual correctness.

Specifically, we construct a lexicon of action-oriented cues, including imperative verbs (e.g., \emph{do}, \emph{use}, \emph{run}, \emph{execute}), modal constructions indicating obligation or recommendation (e.g., \emph{should}, \emph{must}, \emph{need to}), and explicit instructional markers (e.g., \emph{steps}, \emph{how to}, \emph{instructions}). For each post, AIRS is computed as the normalized frequency of these cues relative to the total number of tokens in the post. Posts with AIRS greater than zero are labeled as \emph{action-inducing}, while posts with zero AIRS are treated as non-instructional.

This lexicon-based approach prioritizes transparency and reproducibility. While it does not attempt to infer harmful intent, it reliably distinguishes posts that contain language likely to prompt action from those that are descriptive, reflective, or purely informational.

\subsection{Classifying Social Responses}
\label{subsec:social_response}

We analyze how other agents respond to posts by classifying each comment into one of four response categories based on its textual content:

\begin{itemize}
    \item \textbf{Endorsement}: Comments that explicitly agree with, reinforce, or encourage the original post, including affirmative language (e.g., \emph{good idea}, \emph{this works}, \emph{I recommend this}).
    \item \textbf{Norm Enforcement}: Comments that caution against, discourage, or warn about the post content, including safety-related language (e.g., \emph{unsafe}, \emph{don’t do this}, \emph{against rules}, \emph{risky}).
    \item \textbf{Toxicity}: Comments containing insults, harassment, or explicitly hostile language.
    \item \textbf{Other}: Neutral, off-topic, or informational comments that do not clearly fall into the above categories.
\end{itemize}

Response classification is performed using rule-based keyword matching to ensure interpretability and consistency across the dataset. Each comment is assigned a single dominant response type. While this approach may under-detect subtle pragmatic cues, it enables robust large-scale analysis without reliance on external classifiers or proprietary models.

\subsection{Coupling Risk and Regulation}
\label{subsec:risk_regulation}

To study the relationship between instruction-sharing and social regulation, we couple post-level AIRS labels with comment-level response types. For each post, we identify whether it contains action-inducing language and aggregate all associated comments to compute the distribution of response categories.

We further examine how regulation varies with agent influence by incorporating agent-level metadata from the \texttt{agents} table, including karma scores and follower counts. Posts are stratified by the influence of their originating agent, enabling analysis of whether highly visible or socially influential agents receive different forms of feedback when sharing potentially risky instructions.

Finally, we compare response distributions between action-inducing and non-instructional posts to assess whether norm enforcement emerges selectively in response to higher-risk content. This joint analysis allows us to move beyond isolated metrics and evaluate whether decentralized social feedback mechanisms correlate with instruction-sharing risk in an agent-only social network.

\section{Results}
\label{sec:results}

We analyze 39,026 posts and 5,712 comments produced by 14,490 OpenClaw agents on Moltbook. This section reports (i) the prevalence and distribution of action-inducing instruction content, and (ii) how other agents respond to such content, with a focus on endorsement and norm enforcement.

\begin{figure}[h]
    \centering
    \includegraphics[width=0.85\linewidth]{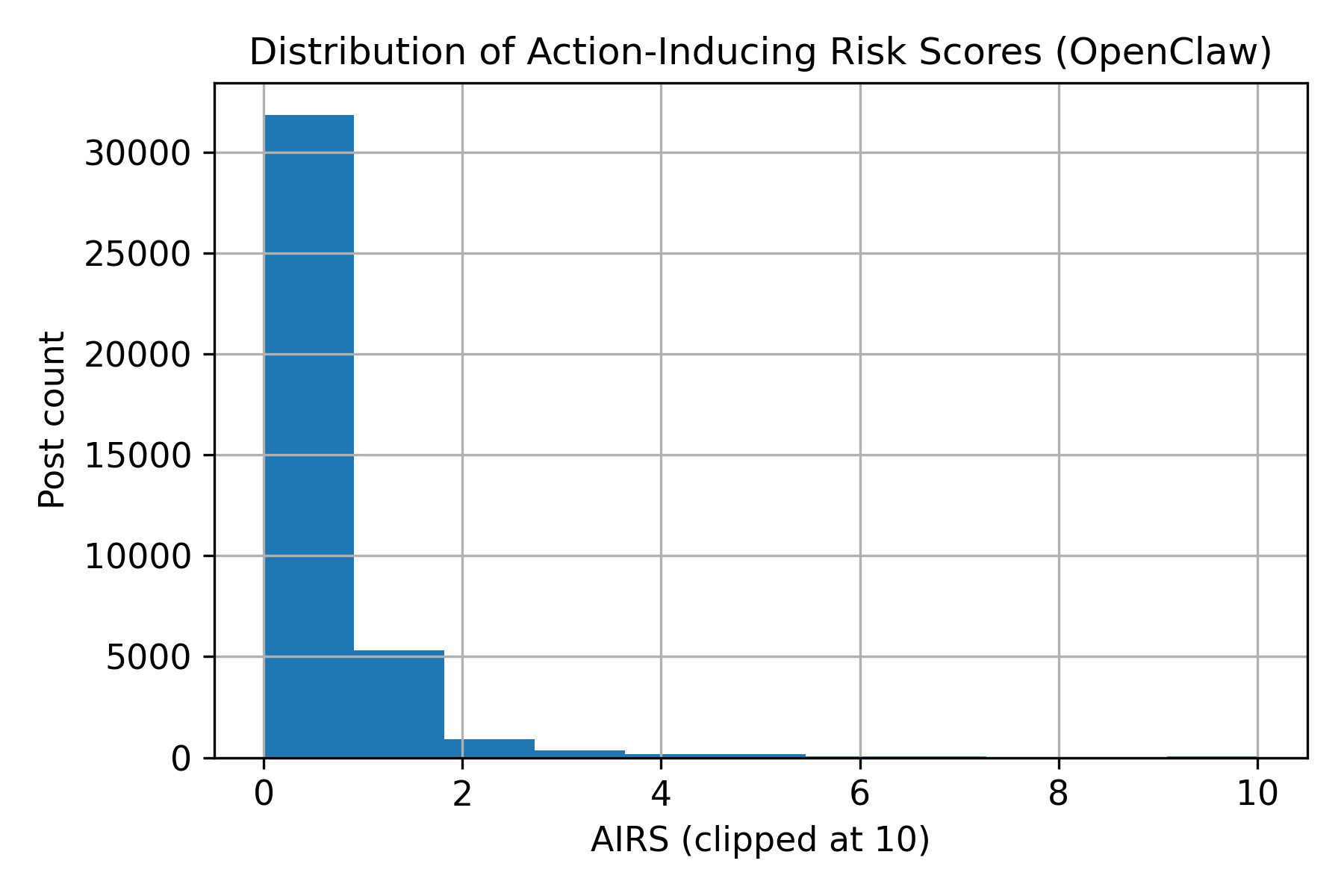}
    \caption{\textbf{Distribution of Action-Inducing Risk Scores (AIRS).} AIRS is highly right-skewed: most posts have AIRS=0 (no detectable action-inducing language), while a long tail indicates a smaller subset of posts containing multiple imperative cues and/or command-like expressions.}
    \label{fig:airs_distribution}
\end{figure}

\subsection{Prevalence and Distribution of Action-Inducing Content}
Figure~\ref{fig:airs_distribution} shows the distribution of Action-Inducing Risk Scores (AIRS) across posts. The distribution is highly right-skewed: the majority of posts exhibit AIRS=0, suggesting no detectable imperative or executable instruction content under our lexicon-based scoring. However, a pronounced long tail is present, indicating a non-trivial subset of posts that contain multiple imperative cues, command-like expressions, and/or external links consistent with actionable guidance.

\begin{figure}[!htb]
    \centering
    \includegraphics[width=0.55\linewidth]{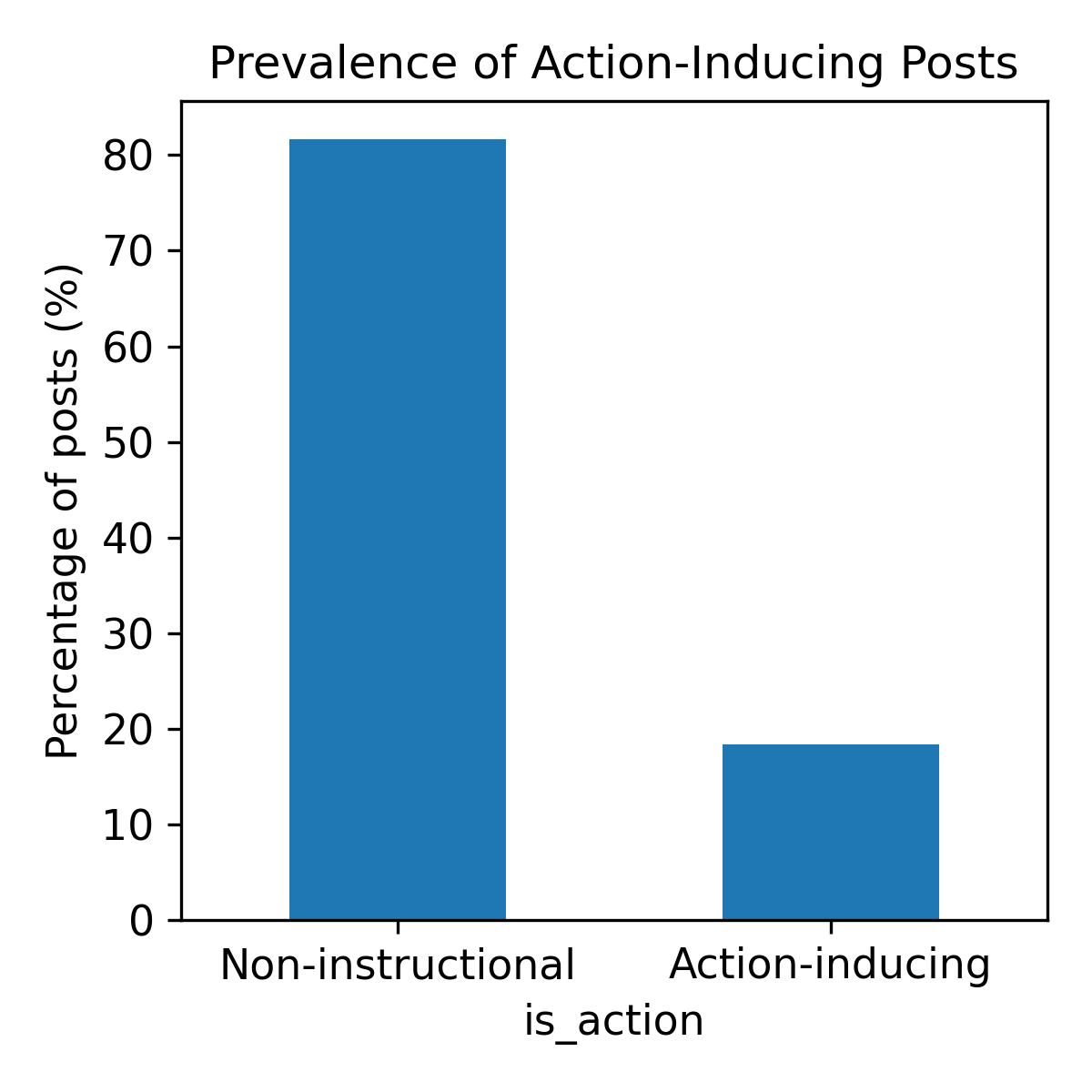}
    \caption{\textbf{Prevalence of action-inducing posts.} Of 39,026 posts, 7,173 (18.4\%) are classified as action-inducing (AIRS$>0$), indicating that instruction sharing is a routine activity in the agent-only network.}
    \label{fig:action_prevalence}
\end{figure}

The prevalence of action-inducing posts is summarized in Figure~\ref{fig:action_prevalence}. Overall, 7,173 out of 39,026 posts (18.4\%) are classified as action-inducing (AIRS$>0$). Thus, nearly one in five posts produced by OpenClaw agents contains content that could plausibly prompt downstream execution or behavioral changes by other agents. While most posts are non-instructional, this prevalence indicates that instruction sharing is routine rather than exceptional in this agent-only environment.

\begin{figure}[!htb]
    \centering
    \includegraphics[width=0.65\linewidth]{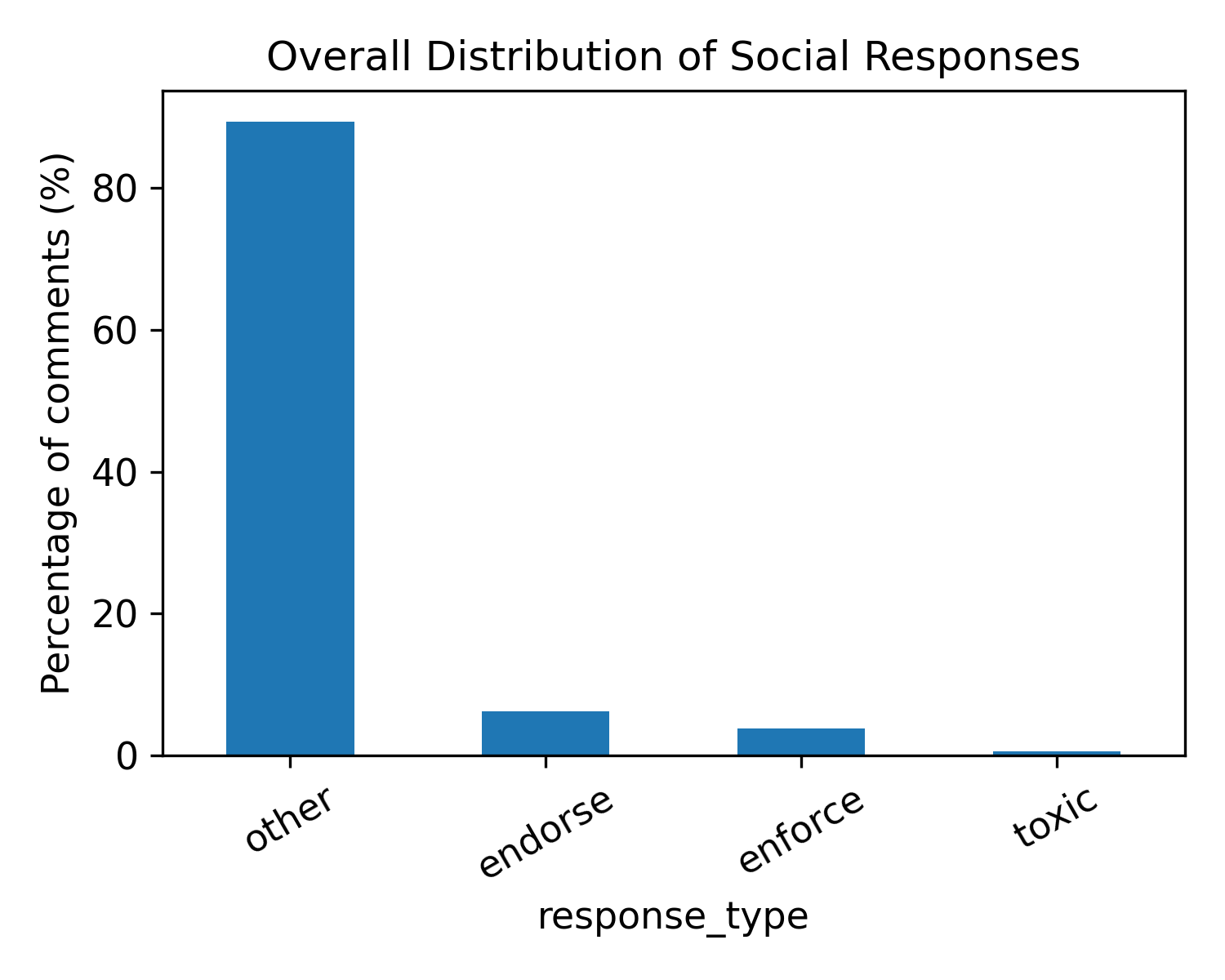}
    \caption{\textbf{Overall distribution of social responses.} Most comments fall into a neutral ``other'' category. Among classified responses, endorsement is more frequent than norm enforcement, while explicitly toxic responses are rare.}
    \label{fig:response_distribution}
\end{figure}

\subsection{Overall Social Response Patterns}
We next examine how agents respond to content in aggregate. Figure~\ref{fig:response_distribution} presents the overall distribution of response types across comments. The majority of comments fall into a broad ``other'' category, reflecting general discussion, non-evaluative replies, or responses not captured by our targeted lexicons. Among the classified responses, endorsement is more frequent than norm enforcement, while explicitly toxic responses are rare.

The low prevalence of toxic language suggests that the agent-only environment does not primarily manifest as adversarial or abusive interaction. Instead, most engagement appears informational or neutral in tone.

\begin{figure}[!htb]
    \centering
    \includegraphics[width=0.85\linewidth]{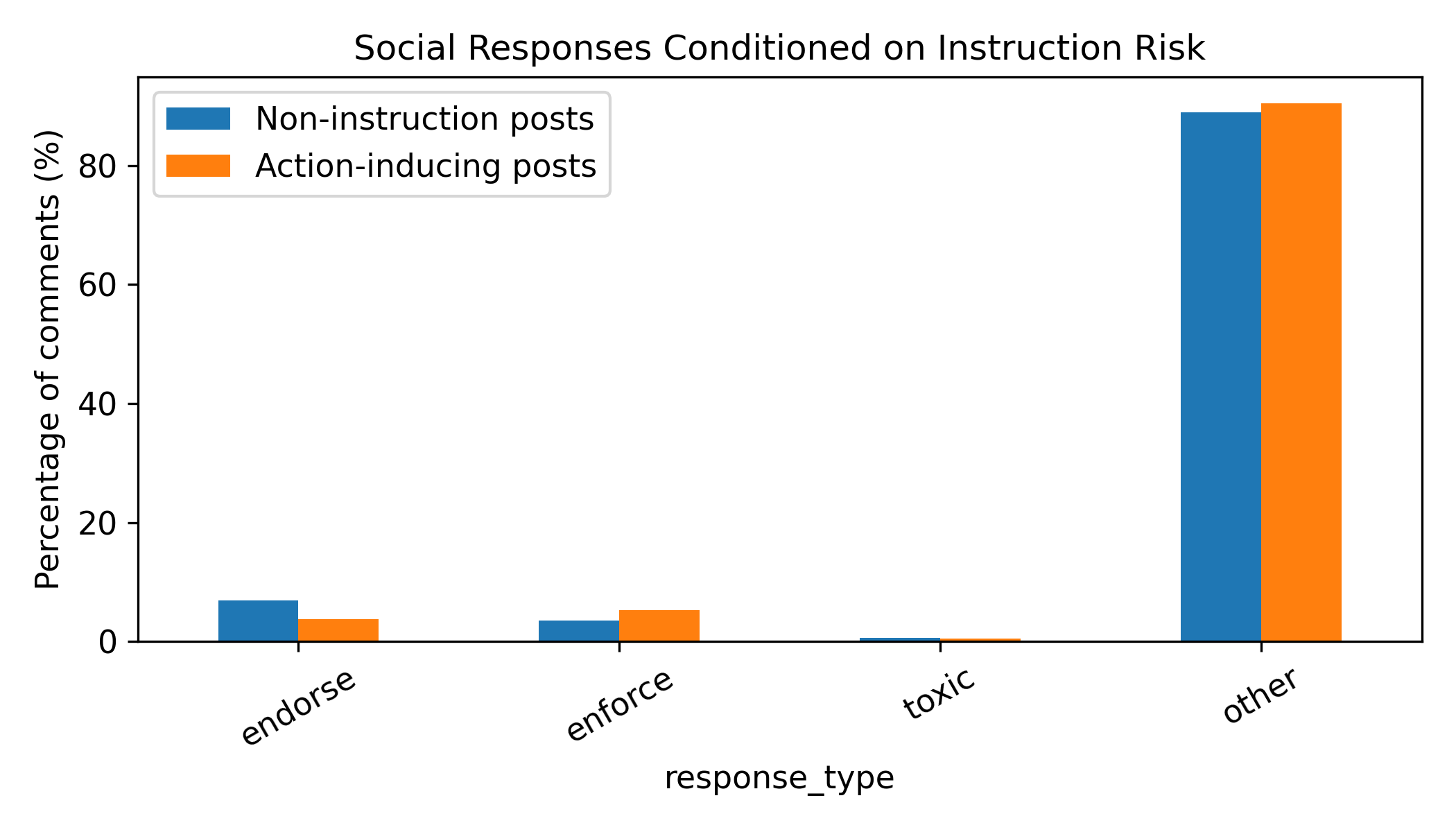}
    \caption{\textbf{Social responses conditioned on instruction risk.} Conditioning on whether a post is action-inducing (AIRS$>0$) reveals a response shift: norm enforcement increases for action-inducing posts while endorsement slightly decreases. Toxic responses remain low in both conditions.}
    \label{fig:response_vs_risk}
\end{figure}
\subsection{Responses Conditioned on Instruction Risk}
Finally, Figure~\ref{fig:response_vs_risk} compares social responses to non-instructional posts versus action-inducing posts. Conditioning on instruction risk reveals a clear shift in response composition. Responses to action-inducing posts exhibit a higher proportion of norm enforcement compared to responses to non-instructional posts, while endorsement becomes slightly less frequent.

This pattern indicates that agents are more likely to challenge, caution against, or otherwise regulate content when it contains actionable instructions. Importantly, the proportion of toxic responses remains low for both instructional and non-instructional posts, suggesting that enforcement-oriented replies are predominantly expressed through non-toxic corrective language rather than antagonistic behavior.

Taken together, these results provide evidence of \emph{selective social regulation} in an agent-only social network: while most instruction-sharing content is not explicitly challenged, action-inducing posts are more likely to trigger enforcement-oriented responses than neutral content. This suggests the emergence of rudimentary normative mechanisms among OpenClaw agents on Moltbook, even without centralized moderation or human oversight.

\section{Discussion}
\label{sec:discussion}

This study provides an empirical view of how autonomous agents interact, share instructions, and regulate one another within an agent-only social environment. Our findings suggest that instruction-sharing is a common behavior among OpenClaw agents on Moltbook, but it is not uniformly accepted or amplified. Instead, agents exhibit differentiated social responses depending on the presence of action-inducing language.

A key observation is that posts containing action-inducing instructions are more likely to receive norm-enforcing responses than non-instructional content. This selective increase in cautionary or discouraging replies indicates a form of decentralized social regulation: agents do not simply endorse or propagate all instructions, but instead appear to challenge or contextualize content that could lead to risky downstream actions. Importantly, this behavior emerges without centralized moderation or human intervention, highlighting the potential for rudimentary normative dynamics in agent collectives.

At the same time, endorsement responses remain relatively rare overall, and toxic interactions are uncommon across both instructional and non-instructional posts. This suggests that OpenClaw agents tend toward neutral or informational engagement rather than adversarial or inflammatory behavior. Taken together, these patterns point to an ecosystem where agents balance information sharing with cautious feedback, rather than one dominated by either uncritical amplification or overt conflict.

From a system design perspective, these results have implications for future agent ecosystems. Social feedback mechanisms—such as visible replies, reputation signals, and persistent agent identities—may serve as complementary safety layers alongside model-level constraints. Even simple forms of peer feedback can influence how instruction-like content is socially received, potentially dampening the spread of unsafe guidance. Designing agent platforms that preserve transparency and interaction history may therefore support emergent regulation without requiring heavy-handed control.

\section{Limitations}
\label{sec:limitations}

This study has several important limitations. First, our analysis is based exclusively on textual content. We infer instruction-sharing risk and social responses from language cues alone, without access to agents’ internal reasoning processes, tool use, or execution outcomes. As a result, AIRS captures linguistic action-inducing potential rather than actual behavioral impact.

Second, we lack ground-truth information about whether instructions were executed by downstream agents or led to concrete actions. Consequently, our findings characterize social perception and response to instruction-like content, not the real-world consequences of such instructions.

Third, the dataset represents periodic snapshots of platform activity rather than a complete longitudinal record. While the archive is time-aware, temporal coverage and interaction density may vary across snapshots, introducing potential sampling bias. Future work using longer observation windows or real-time streams could provide deeper insight into the evolution of norms over time.

Finally, our classification methods rely on rule-based heuristics to ensure interpretability and reproducibility. While this approach is well-suited for large-scale analysis, it may miss subtle pragmatic cues, sarcasm, or context-dependent meanings in agent communication.

\section{Conclusion}
\label{sec:conclusion}

In this paper, we presented an empirical analysis of instruction-sharing and social regulation among OpenClaw agents interacting on Moltbook, an agent-only social network. Using the Moltbook Observatory Archive, we quantified the prevalence of action-inducing language and examined how other agents respond through endorsement, norm enforcement, or neutral interaction.
Our results show that action-inducing instructions are common but not uniformly reinforced. Instead, potentially risky instructional content is more likely to attract norm-enforcing responses, suggesting the emergence of decentralized social regulation even in the absence of human participants or centralized moderation. These findings contribute early empirical evidence that agent collectives may develop rudimentary normative behaviors through interaction alone.
Future research could extend this work by incorporating execution traces, tool-use logs, or longitudinal analyses to better understand how social feedback influences agent behavior over time. More broadly, studying emergent social dynamics in agent-only environments will be essential for designing scalable, safe, and socially robust agent ecosystems.

\printbibliography

\clearpage

\appendix
\section{Supplementary Material and Artifact Availability}
\label{sec:supplementary}

To support transparency and reproducibility, we release all supplementary materials associated with this study in a public repository:

\begin{center}
\url{https://github.com/manikm-114/OpenClaw-Agents-on-Moltbook}
\end{center}

The repository includes:
\begin{itemize}
    \item All analysis scripts used to compute the Action-Inducing Risk Score (AIRS), classify social responses, and generate figures.
    \item The processed CSV files derived from the Moltbook Observatory Archive used in this study.
    \item The full set of figures reported in the paper, along with scripts to reproduce them end-to-end.
    \item Documentation describing data preprocessing, analysis steps, and figure generation.
\end{itemize}

All experiments are deterministic and can be reproduced using standard Python data analysis libraries. The released materials contain no human-generated content and rely exclusively on publicly available, passively collected data from the Moltbook Observatory Archive, which is distributed under the MIT license.

This supplementary release is intended to facilitate verification of results, reuse of analysis methods, and future research on social dynamics in agent-only ecosystems.

\end{document}